\documentclass{v16nufact}

\usepackage{gensymb}
\usepackage{wrapfig}

\def\be{\begin{equation}}
\def\ee{\end{equation}}
\def\bea{\begin{eqnarray}}
\def\eea{\end{eqnarray}}

\newcommand{\Photo}{}

\begin{document}
\vspace*{4cm}
\title{Results and prospects of the latest $\nu_\mu$CC0$\pi$ analyses from the T2K near detectors}

\author{ Stephen Dolan for the T2K Collaboration }

\address{Denys Wilkinson Building, Keble Road, University of Oxford, Oxford, OX1~3RH, UK}

\maketitle\abstracts{
In order to make precision measurements of GeV-scale neutrino oscillations a detailed understanding of both GeV-scale neutrino-nucleon scattering cross sections and nuclear effects are essential. To achieve this, many of the latest analyses at the T2K off-axis near detector (ND280) complement existing results by utilising new techniques to measure muon neutrino charged-current interactions without pions in the final state (CC0$\pi$ interactions). These techniques include water subtraction to obtain a double-differential cross section on oxygen and measurements of differential cross sections using proton kinematics or composite proton-muon variables (such as single transverse kinematic imbalance). In this work these latest CC0$\pi$ measurements are presented with a view to demonstrating their potential to provide complementary precision cross-section measurements and innovative probes of nuclear effects.}

\section{Introduction}\label{sec:intro}

Accelerator-driven long baseline neutrino oscillation experiments rely on the event-by-event reconstruction of neutrino energy from interactions of few-GeV neutrino beams with nuclear targets. The observed final state of such interactions depends on both the interaction mode and various different nuclear effects, including Fermi motion (FM), final state interactions (FSI) and multi-nucleon correlations (np-nh), in addition to the neutrino energy. It is therefore essential to develop a detailed understanding of both interaction mode cross sections and nuclear effects in order to control potential bias in precision measurements of the oscillation parameters. 

The predominant interaction mode of neutrinos with energies relevant to the Tokai to Kamioka (T2K) experiment is charged-current quasi-elastic (CCQE) scatting, in which a neutrino is converted to a charged lepton via the exchange of a W boson in the following reaction: $\nu_\ell n \rightarrow \ell^- p$. This accounts for $\sim$75\% of all events at the T2K peak neutrino beam energy (0.6 GeV). To avoid excessive dependence on neutrino scattering models, CCQE scattering can be probed through measurements of all interactions with a single charged lepton and no pions in the final state to form CC0$\pi$ cross sections. Recent measurements from both the T2K off-axis near detector (ND280)~\cite{T2Kcc0pi} and MINERvA~\cite{minervacc0pi} have measured this cross section for muon neutrinos scattering on hydrocarbon (CH) as a function of outgoing lepton kinematics which, for CCQE scatters free of nuclear effects, fully characterise the interaction. However, the presence of nuclear effects limits the ability of muon kinematics to describe the interaction, since FM and potential np-nh correlations are not accounted for. Nuclear effects also allow other `CCnonQE' interaction modes to exhibit a CC0$\pi$ topology, for example pion-production modes in which the pion is absorbed inside the nuclear medium by FSI processes. Furthermore, the nuclear effects observed in neutrino-hydrocarbon scattering at ND280 are not expected to be the same as for the neutrino-water scattering observed by the Super-Kamiokande far detector, thereby complicating ND280's ability to provide constraints for neutrino oscillation analyses. To better understand neutrino-nucleon scattering and provide useful constraints for oscillation measurement, it is therefore clear that CH target measurements using muon kinematics alone are not enough and that it is essential to develop complimentary measurements, such as the proton kinematics measurement presented in the aforementioned MINERvA analysis~\cite{minervacc0pi}. 

In this work, such complimentary measurements at the T2K off-axis near detector (ND280) are presented. Following a summary of the T2K experiment in section~\ref{sec:t2k}, section~\ref{sec:xsec} demonstrates a measurement of a CC0$\pi$ cross section on water and exhibits three ongoing measurements using the outgoing proton kinematics, in addition to the muon kinematics, to untangle the role of nuclear effects.

\section{The T2K experiment}\label{sec:t2k}

The T2K long baseline neutrino oscillation experiment consists of a muon (or antimuon) neutrino beam produced at the J-PARC facility, a suite of near detectors placed 280 m downstream and the Super-Kamiokande (SK) water Cherenkov far detector located a further 295 km away. Both SK and the ND280 near detector are placed 2.5 $\degree$ off-axis from the beam centre, allowing exposure to a relatively narrow band neutrino flux with a peak energy of 0.6 GeV~\cite{T2K}. The neutrino flux is well constrained by hadron production measurements from the NA61/SHINE experiment~\cite{T2KFlux}.

To measure the neutrino oscillation parameters, T2K observes electron/muon neutrino appearance/disappearance within the beam as a function of reconstructed neutrino energy. The predominant systematic uncertainty on oscillation measurements stems from the variation in the event rate at the far detector due to cross-section and nuclear effect uncertainties (4-6\%)~\cite{T2KOsc}.

\subsection{The ND280 off-axis near detector}\label{sec:nd280}

\begin{wrapfigure}{r}{0.4\textwidth}
\vspace{-9mm}
\includegraphics[scale=0.27]{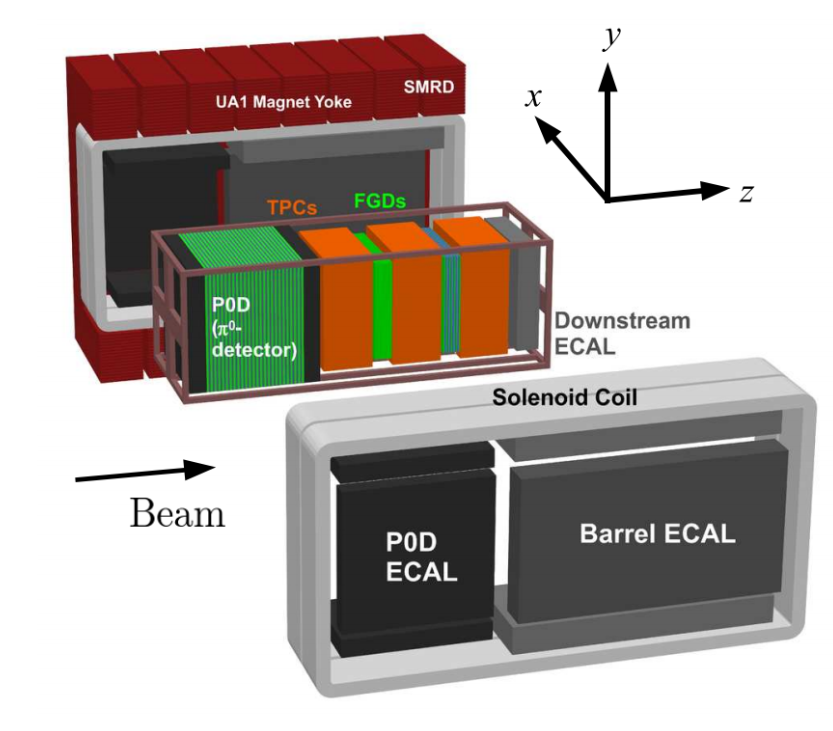}
\caption{ \small An exploded view of the ND280 off-axis near detector labelling each sub-detector.} 
\label{fig:nd280}
\vspace{-1mm}
\end{wrapfigure}

The ND280 off-axis near detector is composed of a series of central specialised sub-detectors surrounded by an electromagnetic calorimeter (ECal) which itself is contained within the UA1 magnet, providing a 0.2 T dipole field~\cite{T2K}. A schematic of ND280 is shown in figure~\ref{fig:nd280}.

The most upstream central sub-detector is the pi-zero detector (P$\O$D), which consists of tracking planes of scintillating bars, interleaved with fillable water bags or lead/brass sheets. Downstream of the P$\O$D is the tracker region, comprising of three time projection chambers (TPCs) and two fine grained detectors (FGD1 and FGD2). The FGDs are both made of finely segmented scintillating bars which provide a neutrino target mass and charged particle tracking, whilst FGD2 also contains inactive water layers. Events leaving the FGDs or the P$\O$D can be tracked in to the TPCs, which provide high-resolution tracking and thereby allow the curvature of charged particles to be used to make excellent measurements of charged particle momentum and identity.

\section{$\nu_\mu$ CC0$\pi$ cross-section measurements} \label{sec:xsec}

All analyses presented here aim to measure a muon neutrino CC0$\pi$ cross section using neutrino interactions in ND280 observed whilst T2K was running with a muon neutrino beam. Each analysis uses events produced using the NEUT~\cite{NEUT} event generator, with final state particles propagated using GEANT 4~\cite{GEANT4} and readout using a detector electronics simulation to validate an event selection aiming to find exactly one $\mu^-$, potentially some nucleons and nothing else. For each analysis the proportion of all signal events selected (efficiency) and the proportion of events selected that are signal (purity) are given as metrics for the effectiveness of the selection.

In each analysis the selection provides events in bins of a reconstructed quantity, so the events must be deconvolved from detector smearing/acceptance in order to produce an experiment independent cross section. This deconvolution is performed using a maximum-likelihood template fitting method or an iterative unfolding approach based on D'Agostini's method~\cite{dagostini}. These are both described and utilised in analysis 1 and 2 of reference~\cite{T2Kcc0pi} respectively. 

Statistical and systematic errors are propagated by re-evaluating the cross section for many Poissonian fluctuations of the selected number of events in each reconstructed bin, and for plausible variations of detector response, flux and neutrino scattering/nuclear models respectively.

\subsection{A double-differential CC0$\pi$ cross section on water}\label{sec:p0dwater}

A measurement of the double-differential CC0$\pi$ cross section on water in muon momentum and angle is made using the P$\O$D by subtracting the event rates measured when the P$\O$D water bags were empty (`water-out') from when they were full ('water-in`). The measurement uses data equivalent to $5.8 \times 10^{20}$ protons on target (POT). 

CC0$\pi$ events are selected by requiring a single track, starting in the P$\O$D fiducial volume, that is matched into the TPC where its charge must be found to be negative and the track must facilitate an accurate momentum measurement. When applied to the simulated events, the water-in/out selection finds 18.2\%/17.5\% efficiency and 77.5\%/82.0\% purity for a signal definition comprising of all CC0$\pi$ events starting in the P$\O$D fiducial volume. The predominant backgrounds are from charged-current events with untagged pions in the final state, 10.6\%/14.1\% from those with a single final state pion (CC1$\pi$) and 4.1\%/2.9\% from those with more (CCMulti$\pi$), which are constrained through specialised sidebands. A CC1$\pi$ sideband selects events with 2 P$\O$D tracks and an associated Michel electron (to identify events with low energy secondary muons, likely from charged pion decays), whilst a CCMulti$\pi$ sideband looks for greater than 2 P$\O$D tracks. A background can then be rescaled in the signal selection based on the discrepancy between simulation and data in the respective sideband selection.

\begin{figure}[h]
\vspace{-6.5mm}
\centering
\includegraphics[width=0.75\textwidth]{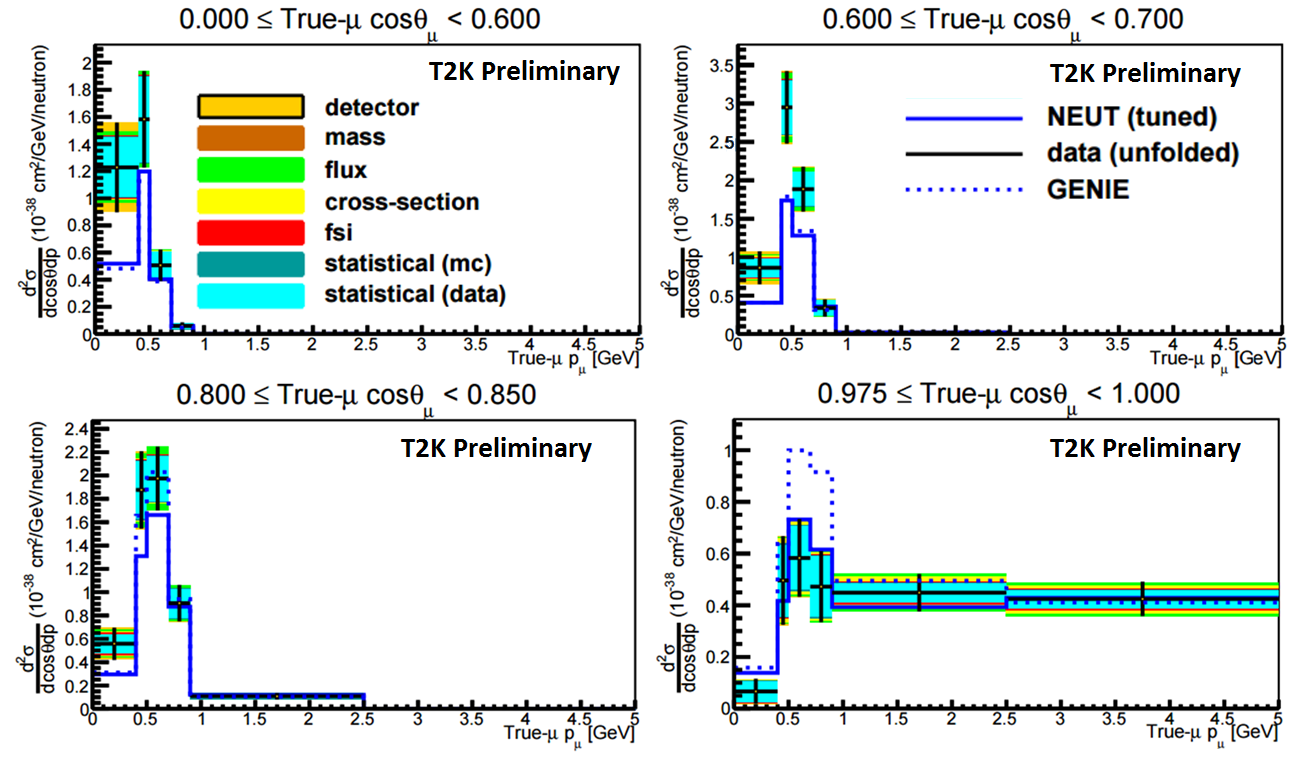}
\caption{ \small The measured double-differential CC0$\pi$ cross section on water compared to the NEUT and GENIE predictions. The errors are shown split by source. Not all angular bins are shown.} 
\label{fig:p0dresult1}
\end{figure}

Selected events are split into bins of the muon momentum and angle and unfolded using the aforementioned D'Agostini-based method with a single iteration to give a double-differential cross section. A sub-sample of the result, complete with statistical and systematic uncertainties, is shown in figure~\ref{fig:p0dresult1}, compared to the predictions of the NEUT and GENIE~\cite{GENIE} event generators. Save for an overestimate by GENIE in the most forward going angular bin, the generators are in good agreement with the data in high momentum, forward going muon bins. However, both generators seem to substantially underestimate the low momentum, high angle muon bins. 

\begin{figure}[h]
\vspace{-1.5mm}
\centering
\includegraphics[width=0.75\textwidth]{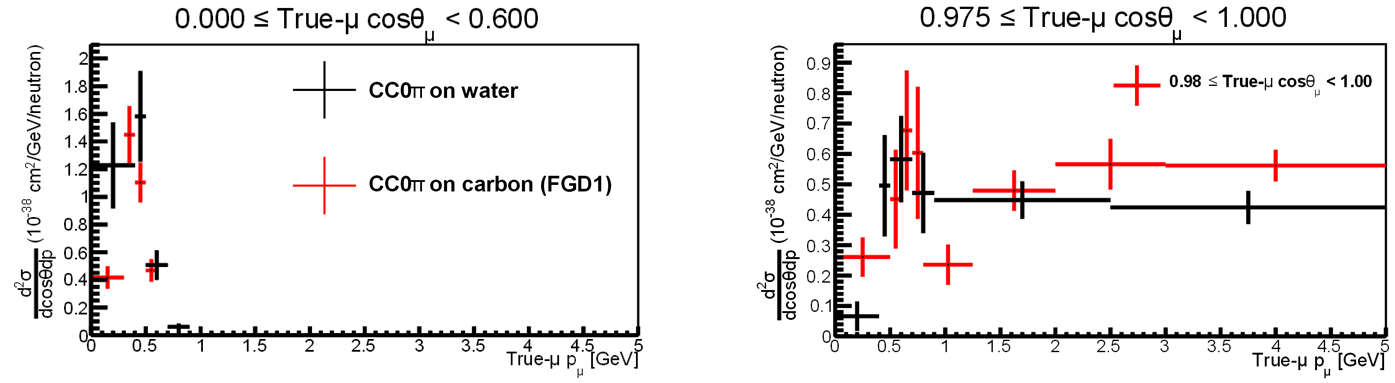}
\caption{ \small The measured double-differential CC0$\pi$ cross section on water compared to a previous result obtained on hydrocarbon, using FGD1. The errors include all uncertainties other than the flux systematic, since these are entirely correlated between the measurements. Not all angular bins are shown and it should be noted that the binning for the two analyses is not identical.} 
\label{fig:p0dresult2}
\end{figure}

By comparing these results to those measured on hydrocarbon~\cite{T2Kcc0pi} a measure of cross section scaling with nucleon number (A-scaling) can be made. This is then a direct probe of nuclear effects on the CC0$\pi$ cross section. An example of such a comparison is shown in figure~\ref{fig:p0dresult2}. Ongoing work will develop ND280's probe of A-scaling by using the FGD2 water layers to measure a CC0$\pi$ cross section ratio between hydrocarbon and water.

\subsection{CC0$\pi$+N$p$ cross sections using proton kinematics}\label{sec:cc0pinp}

As described in section~\ref{sec:intro}, it is difficult to separate the impact of cross sections and nuclear effects using muon kinematics alone. To gain better separation power between various nuclear and cross section effects there are three complimentary analyses at ND280 that aim to measure a CC0$\pi$+N$p$ ($ N\geq1$) cross section as a function of a combination of muon and proton kinematics. Analysis 1 measures a double/triple-differential cross section in muon angle, proton angle and proton momentum. Analysis 2 produces a differential cross section in single transverse kinematic imbalance (STKI), described in~\cite{stv}. Analysis 3 measures a differential cross section as a function of inferred kinematic imbalance (IKI), described below, in bins of muon momentum and angle. All three analyses measure a cross section in a phase space restricted to match ND280 acceptance. Since these analyses have only very recently been applied to real data, they remain under collaboration review and only fake data results, produced using the GENIE event generator with statistics similar to real data ($5.73 \times 10^{20}$ POT), will be shown.

The analyses share a common event selection, requiring events with a vertex in FGD1 and tracks identified as being a muon and proton(s) measured in either FGD1 alone or a combination of FGD1 and the adjacent TPCs. To reject CC1$\pi$ and CCMulti$\pi$ backgrounds a veto on both the presence of Michel electrons and  ECal electromagnetic showers consistent with neutral pion decay $\gamma$s is employed. For a signal definition comprising of all CC0$\pi$+N$p$ events starting in the FGD1 fiducial volume the selection achieves a 12\% efficiency and 84\% purity. Each analysis also utilises sidebands to constrain residual backgrounds. Analyses 1 and 2 select events starting in FGD1 containing a muon and a pion, both tracked into the TPC, to form a  CC1$\pi$ sideband  whilst a muon, a pion and an additional track are required to form a CCMulti$\pi$ sideband. Analysis 3 inverts the Michel electron rejection cut to form a single sideband. 

\begin{figure}[h]
\vspace{-5mm}
\centering
\includegraphics[width=1.0\textwidth]{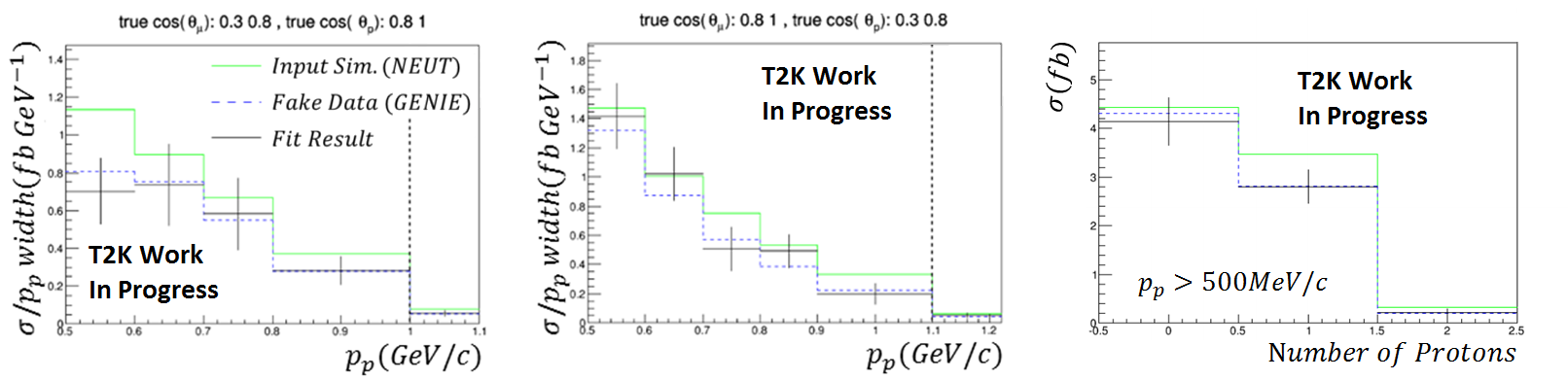}
\caption{ \small The extracted differential CC0$\pi$+N$p$ cross section from fake data produced by the GENIE event generator is shown alongside the input physics simulation from NEUT and the fake data truth. The left and central plot show two bins of a triple-differential cross section in proton momentum, proton angle and muon angle. The last $p_p$ bin (after the dashed line) continues to 30 GeV. The right plot shows the extracted cross section as a function of the multiplicity of protons with at least 500 MeV/c momentum. The error bars for all plots contain systematic and statistical uncertainties. } 
\label{fig:protonresults1}
\end{figure}

Analysis 1 places the selected events into bins of reconstructed muon angle, proton angle and, in bins where statistics allow, proton momentum. A maximum likelihood template fit, described in~\cite{T2Kcc0pi}, is then used to simultaneously extract a CC0$\pi$+N$p$ cross section of events with true proton greater than 500 MeV/c, such that a cross section is only measured for events that ND280 can comfortably reconstruct. As a by-product, a cross section as a function of the multiplicity of protons (above 500 MeV/c momentum) is also obtained; this has the potential to act as an interesting probe of multi-nucleon correlations, which tend to lead to multi-proton final states. A sub-sample of the results after extracting a cross section from fake data are shown in figure~\ref{fig:protonresults1}. The analysis successfully extracts the fake data truth and the difference between the NEUT and GENIE predictions is substantial compared to the scale of the uncertainty, allowing model separation potential. 

\begin{figure}[h] 
\vspace{-1mm}
\centering
\includegraphics[width=1.0\textwidth]{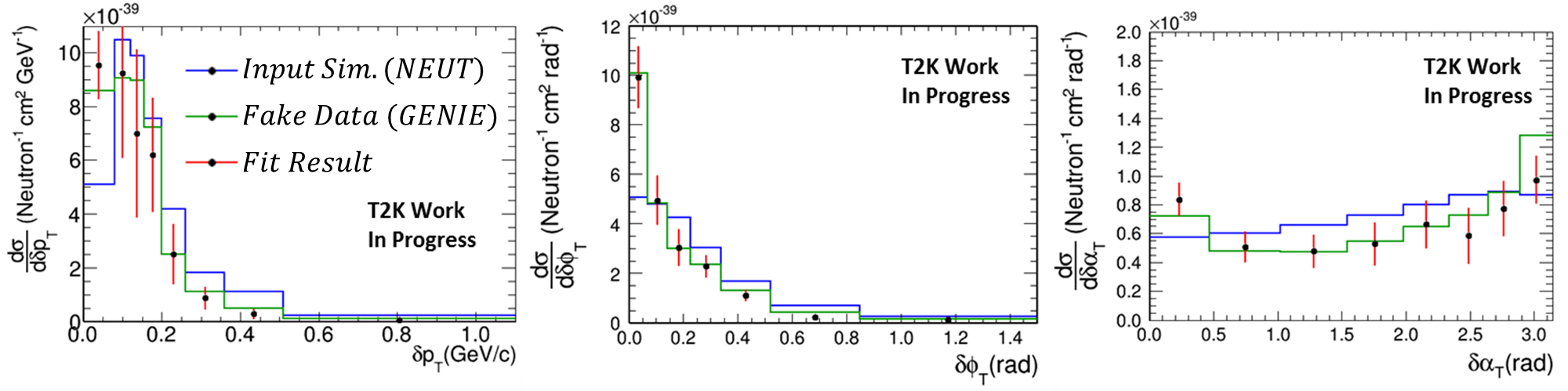}
\caption{ \small The extracted differential CC0$\pi$+N$p$ cross sections from fake data (with the phase space constraints listed in the text) in STKI are shown alongside the input physics simulation from NEUT and the fake data truth from GENIE. The error bars contain combined systematic and statistical uncertainties.}
\label{fig:stikiresults1}
\end{figure}

Analysis 2 places the selected events into bins of the three variables characterising STKI and is described in~\cite{stenpproc,stenuproc}. It has previously been shown that STKI provide an excellent probe of nuclear effects, whilst remaining insensitive to cross section parameter variation, when built from CCQE interactions~\cite{stv}. It has recently been additionally demonstrated that STKI retain much of that sensitivity when built from CC0$\pi$+N$p$ interactions within ND280 acceptance~\cite{stenuproc}. A maximum-likelihood template fit with a Tikhonov regularisation term (to account for the substantial detector smearing of the STKI) is used to build a cross section in bins of STKI with the following constraints on the muon/proton momentum ($p_{\mu / \textrm{p}}$) and angle ($\theta_{\mu / \textrm{p}}$), such that they are measurable at ND280: $p_\mu>250$ MeV/\textit{c}, $p_\textrm{p}>450$ MeV/\textit{c}, $cos(\theta_{\mu})>-0.6$ and $cos(\theta_\textrm{p})>0.4$. The extracted cross sections from fake data are shown in figure~\ref{fig:stikiresults1}. The results indicate a large difference between the NEUT and GENIE prediction demonstrating interesting model separation potential. The difference at large $\delta p_{T}$ and  $\delta \phi_{T}$ stems largely from the generators' very different FSI predictions and the lack of an np-nh contribution in the GENIE version used, whilst the difference at small $\delta p_{T}$ and  $\delta \phi_{T}$ is largely due to the elastic nucleon FSI implementation in the GENIE version used~\cite{stv,lukenpproc}. 

\begin{figure}[h] 
\vspace{-1mm}
\centering
\includegraphics[width=1.0\textwidth]{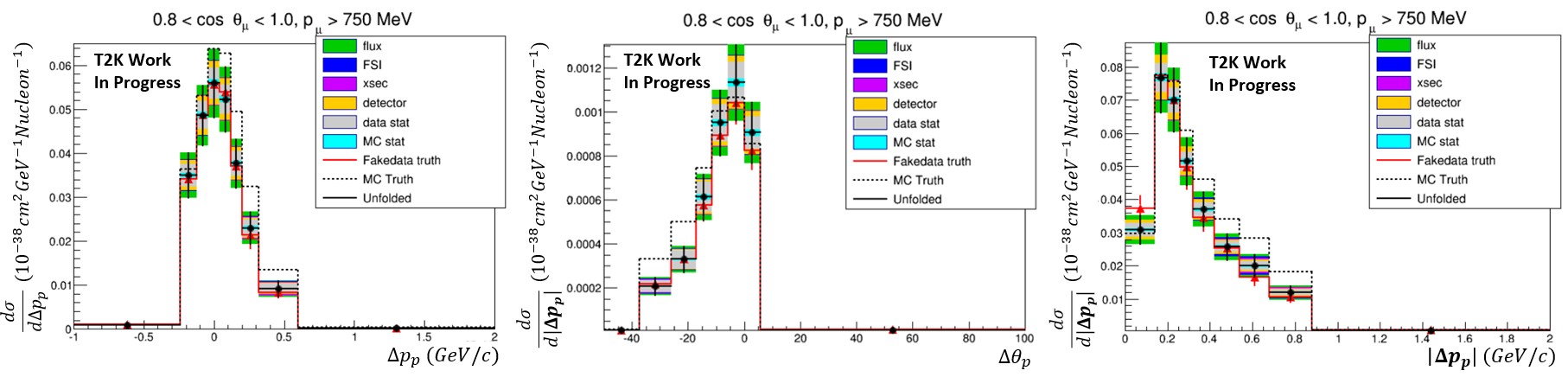}
\caption{ \small The extracted differential CC0$\pi$+N$p$ cross sections from fake data (with the phase space constraints listed in the text) in IKI are shown in select muon kinematic bins alongside the input physics simulation from NEUT and the fake data truth from GENIE. The error bars contain combined systematic and statistical uncertainties.}
\label{fig:ikiresults1}
\vspace{-3mm}
\end{figure}

Analysis 3 exploits IKI. Under the assumption that the incoming neutrino scattered off a stationary neutron, underwent a CCQE interaction and the subsequently ejected proton that did not undergo any FSI processes, the muon kinematics can be used to accurately reconstruct the ejected proton kinematics. These can then be measured and compared to the inference. Any IKI can, on an event by event basis, provide information regarding nuclear effects or the presence of CCnonQE interactions in the selection. The analysis utilises three variables characterising IKI: $\Delta p_p$, the difference between the magnitude of the inferred and measured proton momentum; $\Delta \theta_p$, the difference between the inferred and measured proton angle, and $|{\Delta \mathbf{p_p}}|$, the magnitude of the difference between the inferred and measured proton three momentum. The D'Agostini-based unfolding method is used with a single iteration to measure the three IKI variables in 5 bins of muon momentum and angle, with the following restrictions on proton kinematics, such that the events are measurable at ND280: $p_\textrm{p}>450$ MeV/\textit{c}, $cos(\theta_{\mu})>-0.6$. A subset of the results, along with a full error budget, applied to fake data produced by the GENIE event generator are shown in figure~\ref{fig:ikiresults1}. The results are able to accurately reproduce the fake data with error bars small compared to the NEUT-GENIE model differences, thereby showing model separation potential.

\section{Conclusions}\label{sec:conclusion}

As long baseline neutrino oscillation experiments gather more data, the precision of oscillation analyses become increasingly dependent on an improved understanding of both neutrino-nucleon cross sections and the myriad of nuclear effects related to such scattering. To work towards disentangling these effects, T2K have made the first measurement of the muon-neutrino CC0$\pi$ cross section on water, to compliment the existing measurement on hydrocarbon. Since this measurement probes events on the same target and with the same interaction mode as is measured at SK, it is also particularly useful for providing effective cross section constraints from the near detector to use in oscillation analyses. 

Furthermore, T2K are also measuring the CC0$\pi$+N$p$ cross section on hydrocarbon using proton kinematics, muon kinematics and combinations thereof. These composite measurements are specifically engineered to extract particular nuclear effects whilst remaining insensitive to the free nucleon cross section through the exploitation of STKI and IKI and have been shown to show interesting model separation potential in fake data studies.

\end{document}